\documentstyle[twocolumn,aps,prl]{revtex}
\begin{document}
\input{psfig.sty}
\newcommand{\ketzero}{\mbox{$|0\rangle$}}
\newcommand{\ketone}{\mbox{$|1\rangle$}}
\newcommand{\ketx}{\mbox{$|x\rangle$}}
\title{Approximate quantum counting on an NMR ensemble quantum computer}
\author{J. A. Jones\cite{byline}}
\address{Oxford Centre for Molecular Sciences, New Chemistry Laboratory,
South Parks Road, Oxford OX1 3QT, UK,\\ and Centre for Quantum
Computation, Clarendon Laboratory, Parks Road, Oxford OX1 3PU, UK}
\author{M. Mosca}
\address{Centre for Quantum Computation and Mathematical Institute,
24--29 St Giles', Oxford OX1 3LB, UK}
\date{\today}
\maketitle
\begin{abstract}
We demonstrate the implementation of a quantum algorithm for estimating
the number of matching items in a search operation using a two qubit
nuclear magnetic resonance (NMR) quantum computer.
\end{abstract}

\narrowtext

Quantum computers \cite{Feynman82,Deutsch85} offer the tantalising
prospect of solving computational problems which are intractable for
classical computers.  A variety of algorithms have been developed, most
notably Shor's algorithm for factorising composite numbers in polynomial 
time \cite{Shor94,Ekert96}, and Grover's quantum search algorithm
\cite{Grover96,Grover97}.  Until recently these algorithms were only of 
theoretical interest, as it proved extremely difficult to build a
quantum computer.  In the last few years, however, there has been
substantial progress \cite{Cory96,Cory97,Gersh97} in the construction of
small quantum computers based on nuclear magnetic resonance (NMR)
studies
\cite{Ernst87} of the nuclei of small molecules in solution.  NMR
quantum computers have been used to implement a variety of simple
quantum algorithms, including Deutsch's algorithm
\cite{Jones98,Chuang98} and Grover's algorithm
\cite{Chuang98b,Jones98b,Jones98c}. 

NMR quantum computers differ from other implementations in one important
way: there is not one single quantum computer, but rather a statistical
ensemble of them.  For this reason NMR quantum computers should be
described using density matrices rather than the more usual ket
notation.  In some cases this ensemble nature is irrelevant: it is
possible to prepare the system with an initial density matrix
indistinguishable from that of a pure eigenstate (a pseudo-pure state),
and as long as the result is another pseudo-pure state the behaviour of
an ensemble quantum computer is identical to that of a conventional
quantum computer.  Some algorithms, however, produce a superposition of
states (relative to the natural NMR computational basis) as their final
result, and in such cases the behaviour of an ensemble quantum computer
will be quite different.

An important example is Grover's algorithm when there is more than one
matching item to be found \cite{BBHT96}.  Suppose a search is made over
$N$ items amongst which there are $k$ matching items.  After
$O(\sqrt{N/k})$ evaluations of Grover's search function the quantum
search algorithm will produce an equally weighted superposition of the
$k$ matching items.  With a conventional quantum computer this state
allows any one of the $k$ matching items to be determined at random, as
a measurement will result in one of the states contributing to the
superposition.  With an ensemble quantum computer, however, different
members of the ensemble will result in different states, and the final
observed signal will be an average over the $k$ matching values.  In
general it will be difficult or impossible to deduce anything about
individual matching items from this ensemble average, and so NMR quantum
computers will not be capable of carrying out conventional Grover
searches when more than one item matches the search criteria.

An alternative approach to searching is to count the number of matching
items found in some desired portion of the search space.  Clearly a
bisection search will then permit the first matching item, for example,
to be located in approximately $\log_2(N)$ attempts.  This is only a
sensible strategy if some efficient algorithm for counting matches can
be found.  Fortunately this can be achieved by a simple modification of
Grover's quantum search, approximate quantum counting
\cite{BBHT96,BHT98,Mosca98}.

Suppose we have a function $f(x)$ which maps $n$-bit binary strings to a
single output bit, so that $f(x)=0$ or $1$.  In general there will be
$N=2^n$ possible input values, with $k$ values for which $f(x)=1$.
Grover's quantum search \cite{Grover96,Grover97,BBHT96} allows one of
these $k$ items to be found, while quantum counting
\cite{BBHT96,BHT98,Mosca98} allows the value of $k$ to be estimated.  The 
counting algorithm can be considered as a method for estimating an
eigenvalue of the Grover iterate $G = H U_0 H^{-1} U_{\overline{f}}$,
which forms the basis of the searching algorithm (the operator $H$
corresponds to the $n$-bit Hadamard transform, $U_0$ maps $\ketzero$ to
$-\ketzero$ and leaves the remaining basis states alone, and
$U_{\overline{f}}$ maps $\ketx$ to $(-1)^{f(x)+1}\ketx$).

Starting from the state $|000\ldots 0\rangle\langle 000\ldots 0|$ apply 
the Hadamard operator $H$ to obtain an equally weighted superposition of
all basis states.  For $0<k<N$ we write
\begin{equation}
H|000\ldots 0\rangle =(|\Psi_+\rangle + |\Psi_-\rangle)/\sqrt{2},
\label{eq:H000}
\end{equation}
where $|\Psi_+\rangle$ and $|\Psi_-\rangle$ are eigenvectors of $G$ with
eigenvalues $e^{\pm i\phi_k}$ and $\cos(\phi_k)=1-2k/N$.  For the two
extreme cases, $k=0$ and $k=N$, $H|000\ldots 0\rangle$ is itself an
eigenvector, and we can write $|\Psi_+\rangle=|\Psi_-\rangle=H|000\ldots
0\rangle$, with eigenvalues given by the formulae above.

Eigenvalue estimation is most easily described by considering a register
which begins the calculation in an eigenvector of $G$, say
$|\Psi_+\rangle$.  An additional \emph{control} qubit is needed which
begins in the state $(\ketzero+\ketone)/\sqrt{2}$; this may be obtained
from $\ketzero$ by a Hadamard transform.  The operator $G$ is then
applied to the target register when the control bit is in state
$\ketone$, that is, a controlled-$G$.  The controlled-$G$ produces the
result
\begin{equation}
\frac{1}{\sqrt{2}}(\ketzero+e^{i\phi_k}\ketone)|\Psi_+\rangle,
\end{equation}
or after $r$ repetitions of the controlled-$G$
\begin{equation}
\frac{1}{\sqrt{2}}(\ketzero+e^{ir\phi_k}\ketone)|\Psi_+\rangle.
\end{equation}
Applying a second Hadamard transform to the control qubit gives
\begin{equation}
\left(\frac{1+e^{ir\phi_k}}{2}\ketzero+\frac{1-e^{ir\phi_k}}{2}\ketone\right) 
|\Psi_+\rangle;
\end{equation}
tracing out the target register and expanding the exponential terms
gives for the final sate of the control qubit
\begin{equation}
\rho=\frac{1}{2}\left(
\begin{array}{cc}
        1 + \cos(r\phi_k)    &   i \sin(r\phi_k)  \\
        -i \sin(r\phi_k)     &   1 - \cos(r\phi_k)
\end{array}
\right).
\end{equation}
The same result is obtained if we replace $|\Psi_+\rangle$ with
$|\Psi_-\rangle$, except that the two off diagonal elements are negated.
Thus the same diagonal elements are also obtained from any superposition
or statistical mixture of the two, such as $H|000\ldots 0\rangle$
(equation \ref{eq:H000}).

A variety of different ensemble measurements can be performed to
characterise the final state of the control qubit, but the simplest
approach is to measure the expectation value of $\sigma_z$.  This
corresponds to determining the population difference between the
$|0\rangle\langle 0|$ and $|1\rangle\langle 1|$ states, and is
proportional to $\cos(r\phi_k)$.  Note that in this case ensemble
quantum computers have an an advantage: with a single quantum computer
it would be necessary to repeat the calculation several times in order
to obtain a statistical estimate of $\cos(r\phi_k)$.

$\phi_k$ can be estimated by varying $r$ (the number of repetitions of
the controlled-$G$).  Estimating $\phi_k$ with sufficient accuracy to
determine $k$ requires roughly $\sqrt{k(N-k)}$ applications of
$G$\cite{Beals98}, while a classical algorithm would require $N$
evaluations of $f$.  It is also possible to estimate $k$ to some desired
accuracy: to obtain an estimate $\tilde{k}$ with accuracy $\epsilon$,
that is
\begin{equation}
| \tilde{k} -k | \leq \epsilon k
\end{equation}
requires on the order of $(1/\epsilon)\sqrt{N/(k+1)}$ applications of $G$
\cite{BHT98,Mosca98,footnote}, while a classical algorithm requires about
$(1/\epsilon^2)N/(k+1)$ evaluations of $f$.

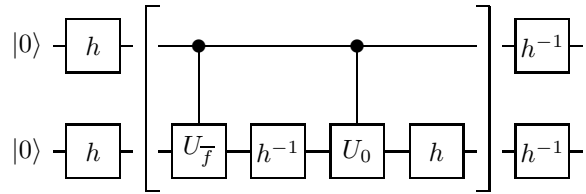
\begin{figure}
\begin{center}
\begin{picture}(220,90)(0,0)
\put(5,60){\makebox(0,0){\ketzero}}
\put(15,60){\line(1,0){5}}
\put(20,50){\framebox(20,20){$h$}}
\put(40,60){\line(1,0){5}}
\put(55,60){\line(1,0){120}}
\put(70,60){\circle*{5}}
\put(70,60){\line(0,-1){30}}
\put(130,60){\circle*{5}}
\put(130,60){\line(0,-1){30}}
\put(185,60){\line(1,0){5}}
\put(190,50){\framebox(20,20){$h^{-1}$}}
\put(210,60){\line(1,0){5}}
\put(5,20){\makebox(0,0){\ketzero}}
\put(15,20){\line(1,0){5}}
\put(20,10){\framebox(20,20){$h$}}
\put(40,20){\line(1,0){5}}
\put(55,20){\line(1,0){5}}
\put(60,10){\framebox(20,20){$U_{\overline{f}}$}}
\put(80,20){\line(1,0){10}}
\put(90,10){\framebox(20,20){$h^{-1}$}}
\put(110,20){\line(1,0){10}}
\put(120,10){\framebox(20,20){$U_0$}}
\put(140,20){\line(1,0){10}}
\put(150,10){\framebox(20,20){$h$}}
\put(170,20){\line(1,0){5}}
\put(185,20){\line(1,0){5}}
\put(190,10){\framebox(20,20){$h^{-1}$}}
\put(210,20){\line(1,0){5}}
\put(50,5){\line(1,0){5}}
\put(50,5){\line(0,1){70}}
\put(50,75){\line(1,0){5}}
\put(180,5){\line(-1,0){5}}
\put(180,5){\line(0,1){70}}
\put(180,75){\line(-1,0){5}}
\end{picture}
\end{center}
\caption{A quantum circuit for implementing quantum counting on a two
qubit NMR quantum computer; the central sequence of gates, surrounded
by brackets, is applied $r$ times.  The upper line corresponds to the
control bit, while the lower line corresponds to the target bit.  A
similar circuit can be constructed for a larger search space by
replacing the target bit by a register and replacing gates applied to
the target by multi-bit versions.  Gates marked $h$ implement the NMR
pseudo-Hadamard operation, while those marked $h^{-1}$ implement the
inverse operation.  Controlled gates are marked by a circle and a
vertical ``control line''.}
\label{fig:circuit}
\end{figure}

A quantum circuit for implementing this algorithm on a two qubit NMR
quantum computer is shown in figure~\ref{fig:circuit}.  This differs
from the conventional circuit in two ways.  Firstly pairs of Hadamard
gates are replaced by an NMR pseudo-Hadamard gate (a $90^\circ_y$
rotation) and its inverse \cite{Jones98}.  Secondly the
controlled-Hadamard gates inside the controlled-$G$ propagator have been
replaced by uncontrolled gates; this is permitted as the intervening
$U_0$ gate has no effect when the control spin is in state $\ketzero$.
This circuit can be used to count the number of solutions to $f(x)=1$
over a one bit search space, but similar circuits exist for larger
search spaces.

This algorithm was implemented using our two-qubit NMR quantum computer
\cite{Jones98}, which uses two ${}^{1}\rm H$ nuclei in a solution of the 
small molecule cytosine in $\rm D_2O$.  All NMR experiments were carried
out on a home-built spectrometer at the Oxford Centre for Molecular
Sciences, with a ${}^{1}\rm H$ operating frequency of $\rm 500\,MHz$.
The two spin-states of the ${}^{1}\rm H$ nuclei act as qubits, and it is
necessary to address each spin individually.  Previous experiments on
this system \cite{Jones98,Jones98b} have used soft pulses to achieve
selective excitation, and errors in these pulses have resulted in
significant distortions in observed spectra.  For these experiments a
different approach was adopted, using non-selective hard pulses whenever 
possible.

\begin{figure}
\begin{center}
\begin{picture}(140,40)(0,0)
\put(0,10){\line(1,0){140}}
\put(10,20){\makebox(0,0)[b]{$\delta$}}
\put(20,10){\line(0,1){20}}
\put(20,30){\line(1,0){20}}
\put(40,30){\line(0,-1){20}}
\put(30,20){\makebox(0,0)[b]{$x$}}
\put(30,35){\makebox(0,0)[b]{$180$}}
\put(70,20){\makebox(0,0)[b]{$\delta+\epsilon_{270}+\delta$}}
\put(100,10){\line(0,1){20}}
\put(100,30){\line(1,0){20}}
\put(120,30){\line(0,-1){20}}
\put(110,20){\makebox(0,0)[b]{$x$}}
\put(110,35){\makebox(0,0)[b]{$180$}}
\put(130,20){\makebox(0,0)[b]{$\delta$}}
\end{picture}
\end{center}
\caption{A pulse sequence implementing a
controlled-$U_{\overline{f_{01}}}$ gate using only hard pulses and 
periods of free precession.  Pulse rotation angles (in degrees) are
marked above each pulse, while pulse phases are marked within a pulse.
Other periods correspond to free precession under the Hamiltonian $\cal
H$ for the time indicated.  These times are chosen such that
$4\delta+\epsilon_{270}=1/(2J_{IS})$ and $\epsilon_{270}=3\pi/\omega$.}
\label{fig:Ufbar01}
\end{figure}
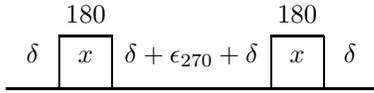

The ${}^{1}\rm H$ transmitter frequency was set in the centre of the
spectrum, so that the two spins have angular frequencies in the rotating
frame of $\pm\omega/2$.  The Hamiltonian can then be written in product
operator notation \cite{Sorensen} as
\begin{equation}
{\cal H}=\frac{\omega}{2}I_z-\frac{\omega}{2}S_z+\pi J_{IS}\,2 I_z S_z
\end{equation}
where $J_{IS}$ is the spin-spin coupling constant, and weak coupling has
been assumed (i.e., $\omega \gg J_{IS}$).  Using a combination of
non-selective pulses and carefully chosen periods of free evolution under
$\cal H$ it is possible to implement many of the necessary gates without
the use of selective pulses.  For example the
controlled-$U_{\overline{f_{01}}}$ gate, which implements the function
when $f(0)=0$ and $f(1)=1$, can be constructed using the pulse sequence
shown in figure~\ref{fig:Ufbar01}.

\begin{figure}
\begin{center}
\begin{picture}(10,90)(0,0)
\put(5,80){\makebox(0,0)[b]{(a)}}
\end{picture}
\begin{picture}(110,90)(0,0)
\put(0,50){\line(1,0){110}}
\put(5,55){$I$}
\put(20,50){\line(0,1){20}}
\put(20,70){\line(1,0){20}}
\put(40,70){\line(0,-1){20}}
\put(30,60){\makebox(0,0)[b]{$y$}}
\put(30,75){\makebox(0,0)[b]{$45$}}
\put(40,70){\line(1,0){20}}
\put(60,70){\line(0,-1){20}}
\put(50,60){\makebox(0,0)[b]{$x$}}
\put(50,75){\makebox(0,0)[b]{$90$}}
\put(70,60){\makebox(0,0)[b]{$\epsilon_{45}$}}
\put(80,50){\line(0,1){20}}
\put(80,70){\line(1,0){20}}
\put(100,70){\line(0,-1){20}}
\put(90,60){\makebox(0,0)[b]{$-x$}}
\put(90,75){\makebox(0,0)[b]{$90$}}
\put(0,10){\line(1,0){110}}
\put(5,15){$S$}
\put(20,10){\line(0,1){20}}
\put(20,30){\line(1,0){20}}
\put(40,30){\line(0,-1){20}}
\put(30,20){\makebox(0,0)[b]{$y$}}
\put(30,35){\makebox(0,0)[b]{$45$}}
\put(40,30){\line(1,0){20}}
\put(60,30){\line(0,-1){20}}
\put(50,20){\makebox(0,0)[b]{$x$}}
\put(50,35){\makebox(0,0)[b]{$90$}}
\put(70,20){\makebox(0,0)[b]{$\epsilon_{45}$}}
\put(80,10){\line(0,1){20}}
\put(80,30){\line(1,0){20}}
\put(100,30){\line(0,-1){20}}
\put(90,20){\makebox(0,0)[b]{$-x$}}
\put(90,35){\makebox(0,0)[b]{$90$}}
\end{picture}
\begin{picture}(10,90)(0,0)
\put(5,80){\makebox(0,0)[b]{(b)}}
\end{picture}
\begin{picture}(50,90)(0,0)
\put(0,50){\line(1,0){50}}
\put(5,55){$I$}
\put(20,50){\line(0,1){20}}
\put(20,70){\line(1,0){20}}
\put(40,70){\line(0,-1){20}}
\put(30,60){\makebox(0,0)[b]{$y$}}
\put(30,75){\makebox(0,0)[b]{$90$}}
\put(0,10){\line(1,0){50}}
\put(5,15){$S$}
\end{picture}
\end{center}
\caption{The sequence of hard pulses and delays shown in (a) is
equivalent to the single selective pulse (b); other selective pulses can 
be implemented in a similar fashion.  Note that the small spin-spin
coupling may be neglected during the short period $\epsilon_{45}$.}
\label{fig:selective}
\end{figure}
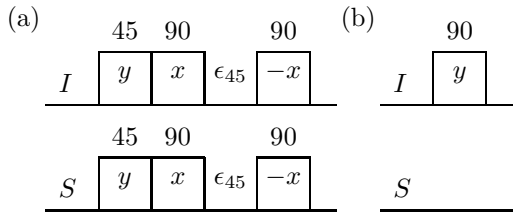

Some gates, however, cannot be implemented without using selective
pulses; for example the pseudo-Hadamard gates within the controlled-$G$
should only be applied to the target spin.  Fortunately it is possible
to create selective pulses using only hard pulses and delays, and this
process is particularly simple when only two spins are involved.  For
short periods of evolution under $\cal H$ the small spin-spin coupling
term can be neglected, and ${\cal H}\approx (\omega/2)(I_z-S_z)$.  Thus
after a time $\epsilon_{45}=\pi/(2\omega)$ the two spins will have
undergone rotations of $\pm 45^\circ$ about their respective $z$-axes.
This $\pm z$-rotation can be converted to a $\pm y$-rotation by
sandwiching the $\tau$ period between $90^\circ_x$ and $90^\circ_{-x}$
pulses (a variant of the more traditional composite $z$-pulse
\cite{Ernst87}).  Combining this with a $45^\circ$ pulse along the
$y$-axis gives an overall $90^\circ_y$ rotation for the first spin
($I$), but no nett rotation for the second spin ($S$), as shown in
figure~\ref{fig:selective}.

With minor variations this approach can be
used to generate selective pulses along any desired axis, and which
excite either $I$ or $S$ as desired.  These selective pulses can then be
used to implement the remaining gates: for example a
controlled-$U_{\overline{f_{10}}}$ can be implemented using the circuit
for controlled-$U_{\overline{f_{01}}}$ with a selective $180^\circ$
pulse applied immediately before and after the other pulses.

The circuit shown in figure~\ref{fig:circuit} encodes the result of the
calculation in the state of the control qubit.  This state could be
characterised in a variety of ways, of which the simplest is to measure
the expectation value of $\sigma_z$ for the spin.  This cannot be
achieved directly, as $z$-magnetisation is not a direct NMR observable,
but an equivalent measurement can be easily made by exciting the spin
with a $90^\circ_y$ pulse and then observing the resulting NMR spectrum.
After appropriate phase correction the integrated intensity of the
corresponding signal gives the desired result.  The phase correction
step requires a reference spectrum \cite{Jones98,Jones98b}, but this is
easily obtained by acquiring a spectrum with $r=0$.

Immediately prior to acquisition a short magnetic field gradient pulse
was applied to destroy the homogeneity of the main field.  This has the
effect of dephasing (and thus rendering undetectable) all off-diagonal
terms in the final density matrix \cite{Jones98b}, with the exception of 
those corresponding to zero quantum coherence \cite{Ernst87}.  The zero
quantum terms can also be removed using the fact that they evolve at
frequencies of $\pm\omega$ under the Hamiltonian $\cal H$.  This zero
quantum filter is easily combined with a standard four-step {\sc
cyclops} phase-cycle \cite{Ernst87}, to reduce instrumental
imperfections.

\begin{figure}
\psfig{file=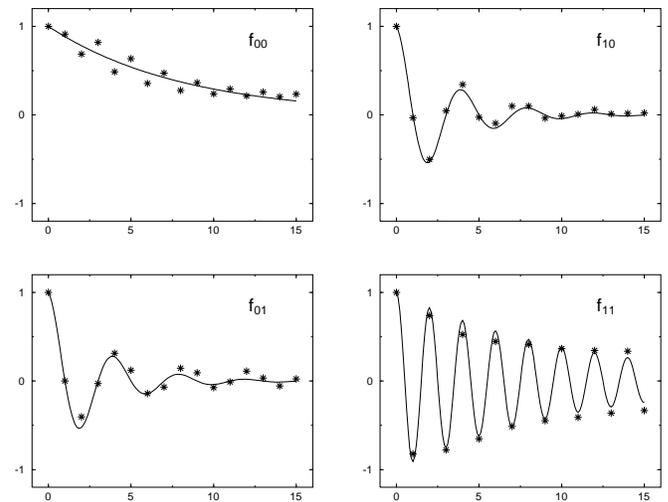,width=87mm}
\caption{Experimental results from our NMR quantum computer for each of
the four possible functions, $f$.  The observed signal intensity is
plotted as a function of $r$, the number of times the controlled-$G$
operator is applied, and all intensities are normalised relative to the
case of $r=0$.  The solid lines are exponentially damped cosinusoids
with the theoretically predicted frequencies, and are plotted merely to
guide the eye.}
\label{fig:results}
\end{figure}

The results of our NMR experiments are shown in
figure~\ref{fig:results}.  Measurements were made for each of the four
possible functions: $f_{00}$ ($k=0$), $f_{01}$ ($k=1$), $f_{10}$
($k=1$), and $f_{11}$ ($k=2$).  In each case the predicted signal is a
cosinusoidal modulation of the signal intensity as a function of $r$,
the number of times the controlled-$G$ is applied, where the frequency
of the modulation, $\phi_k$, depends on $k/N$.  For the two-qubit case,
where $N=2$, the behaviour is particularly simple, with modulation
frequencies of $0$ ($k=0$), $\pi/2$ ($k=1$), and $\pi$ ($k=2$).  In this 
case it is possible to determine $k$ using just {\em one} experiment, with
$r=1$, but spectra were also acquired with larger values of $r$, both to 
demonstrate the principle involved and to explore the build up of errors 
in the calculation.

The experimental results do indeed show a cosinusoidal modulation as
expected, but they deviate from the simple predictions above in a number
of ways.  Firstly all the signals show a clear decay in signal intensity
as $r$ is increased, and this decay is most rapid for $f_{01}$ and
$f_{10}$ (where $k=1$), and least rapid for $f_{11}$ (where $k=2$).  The
simplest explanation for this observation is decoherence: for large
values of $r$ the total length of the pulse sequence is comparable to
the spin-spin relaxation time, $T_2$.  Another likely cause is
imperfections in the pulses applied, in particular those arising from
variations in the strength of the resonant RF field across the sample
($B_1$ inhomogeneity).  Both effects are expected to be most severe when
$k=1$, as these cases have complex $U_{\overline{f}}$ gates which take a
long time to implement, and least severe when $k=2$, in which case
$U_{\overline{f}}$ is just the identity operation.

In addition to the main exponential decay other deviations from the
simple behaviour predicted by theory can be seen.  These effects are
clearest for $f_{00}$ ($k=0$), where alternate signal intensities are
seen to lie alternately above and below the main curve.  Such effects
could in principle arise from many different causes, but numerical
simulations indicate that the major cause is off-resonance effects.
These occur because the applied RF field is not perfectly resonant with
the NMR transitions, but instead is applied a small distance
($\pm\omega/2$) away.  Thus the effect of the field (in the rotating
frame) is not simply to cause a rotation around itself, but rather to
cause a rotation around a tilted axis \cite{Ernst87}.  We are currently
seeking ways to reduce the size of such effects.

Despite these small errors the results are remarkably good, especially
for the case of $f_{11}$.  In this case the experiments have been
repeated with much larger values of $r$, and the cosinusoidal variation
remains clearly visible after 60 or more iterations (data not shown).
Thus our NMR quantum computer is capable of demonstrating quantum
algorithms involving several hundred quantum gates.

\acknowledgments
We thank S. C. Wimperis for helpful discussions.  JAJ thanks C. M. Dobson
for his encouragement.  The OCMS is supported by the UK EPSRC,
BBSRC and MRC.  MM thanks CESG (UK) for their support.


\begin{references}
\bibitem[*]{byline}  To whom correspondence should be addressed at the
New Chemistry Laboratory.
\bibitem{footnote} Note that exact quantum counting gives a roughly
square-root speed up for small values of $k$, while approximate quantum
counting gives a square root speed up in all cases.  Similar results
hold for related problems, as discussed in \cite{Grover98}, such as the
threshold problem\cite{Beals98} and parity determination\cite{Farhi98}.
\end{references}
\end{document}